\documentclass[3p,10pt,a4paper,twoside,fleqn,sort&compress]{filomat}
\usepackage{amssymb,amsmath,latexsym}
\usepackage[varg]{pxfonts}
\usepackage{verbatim}
\usepackage{amsfonts}
\usepackage[colorlinks=true, linkcolor=blue, bookmarks=true]{hyperref}
\newcommand{\be}{\begin{equation}}
\newcommand{\ee}{\end{equation}}
\newcommand{\bea}{\begin{eqnarray}}
\newcommand{\eea}{\end{eqnarray}}
\newcommand{\pd}{\partial}

\newcommand{\Tc}{\mathcal{T}}
\newcommand{\Hc}{\mathcal{H}}
\newcommand{\Mc}{\mathcal{M}}
\newcommand{\Gc}{\mathcal{G}}
\renewcommand{\Re}{\mathrm{Re \,}}
\newcommand{\const}{\mathrm{const}}

\newcommand{\dx}{d^4x}

\newcommand{\nmu}{\nabla_{\mu}}
\newcommand{\nnu}{\nabla_{\nu}}

\newcommand{\FF}{\mathcal{F}}
\newcommand{\GG}{\mathcal{G}}

\newcommand{\arXiv}[1]{\href{http://www.arXiv.org/abs/#1}{#1}}

\newcommand{\FilVol}{xx}
\newcommand{\FilYear}{yyyy}
\newcommand{\FilPages}{zzz--zzz}
\newcommand{\FilDOI}{(will be added later)}
\newcommand{\FilReceived}{Received: dd Month yyyy; Accepted: dd Month yyyy}

\begin{document}

\title{Cosmology of non-local $f(R)$ gravity}

\author[af1]{Ivan Dimitrijevic}
\ead{ivand@matf.bg.ac.rs}
\author[af2,af3]{Branko Dragovich}
\ead{dragovich@ipb.ac.rs}
\author[af4]{Alexey S. Koshelev}
\ead{alexey@ubi.pt}
\author[af1]{Zoran Rakic}
\ead{zrakic@matf.bg.ac.rs}
\author[af5]{Jelena Stankovic}
\ead{jelenagg@gmail.com}
\address[af1]{Faculty of Mathematics, University of Belgrade,Studentski trg 16, 11000, Belgrade, Serbia}
\address[af2]{Institute of Physics, University of Belgrade,Pregrevica 118, 11080 Belgrade, Serbia}
\address[af3]{Mathematical Institute of Serbian Academy of Sciences and Arts, Kneza Mihaila 36, Belgrade, Serbia}
\address[af4]{Departamento  de F\'isica and Centro  de  Matem\'atica e Aplica\c c\~oes,  Universidade  da  Beira  Interior,  6200  Covilh\~a, Portugal}
\address[af5]{Teacher Education Faculty, University of Belgrade, Kraljice Natalije 43,Belgrade, Serbia}
\newcommand{\AuthorNames}{Ivan Dimitrijevic et al.}

\newcommand{\FilMSC}{Primary 83Dxx, 83Fxx; Secondary 83C15}
\newcommand{\FilKeywords}{(nonlocal modified gravity, bounce cosmological solutions, cosmological perturbations, no-ghost conditions)}
\newcommand{\FilCommunicated}{I. Dimitrijevic}
\newcommand{\FilSupport}{This paper was supported in part by  Ministry of Education, Science and Technological Development of the Republic of Serbia, grant No 174012.  AK is supported FCT  Portugal  investigator project IF/01607/2015, by the grant UID/MAT/00212/2013 and  COST  Action  CA15117  (CANTATA).}

\begin{abstract}
We consider a modification of GR with a special type of
a non-local $f(R)$. The structure of the non-local operators is motivated by
the string field theory and $p$-adic string theory. The spectrum is derived explicitly and the ghost-free condition for the model is formulated. We pay special attention
to the classical stability of the de Sitter solution in our model and formulate the
conditions on the model parameters to have a stable configuration.
Relevance of unstable configurations for the description of the
de Sitter phase during inflation is specifically discussed.
\end{abstract}

\maketitle

\makeatletter
\renewcommand\@makefnmark%
{\mbox{\textsuperscript{\normalfont\@thefnmark)}}}
\makeatother

\section{Introduction}

In 2015, the gravity community  celebrated the first century of the General
Relativity (GR), which is viewed as one of the most beautiful
and profound physical theories \cite{wald}. GR is still an acting theory of
gravity usually  presented by Einstein's equation of motion for the
gravitational (metric) field $g_{\mu\nu}$:
\begin{align} R_{\mu\nu} - \frac{1}{2} R g_{\mu\nu} = {8 \pi G} T_{\mu\nu} , \label{eq:1.1}
\end{align}
where $R_{\mu\nu}$ is the
Ricci tensor, $R$ is the Ricci scalar, $T_{\mu\nu}$ is the
energy-momentum tensor of matter, $G$ is the Newtonian constant and the speed
of light is taken $c = 1$. This
Einstein's equation
can be derived from the Einstein-Hilbert action
\begin{align} S = \frac{1}{16\pi G} \int \sqrt{-g}\, R \, d^4x + \int \sqrt{-g} \mathcal{L}_m \, d^4x ,  \label{eq:1.2}
\end{align}
where $g = \det(g_{\mu\nu})$ and $\mathcal{L}_m$ is the Lagrangian of matter.
In this paper we use $(-+++)$ metric signature and stick to $4$ dimensions.

GR has been well tested and confirmed in the Solar system, and
it serves as a theoretical laboratory for gravitational investigations at other
spacetime scales. It has important astrophysical implications predicting
existence of black holes, gravitational lensing and gravitational waves. In
cosmology, GR predicts existence of about $95 \%$
of additional matter, which makes dark side of the Universe. Namely, if GR is
the gravity theory for the Universe as a whole and if the Universe has the
Friedmann-Lema\^{\i}tre-Robertson-Walker (FLRW) metric (which is homogeneous
and isotropic) at the cosmic scale, then there is about $68\%$ of {\it dark
energy}, $27\%$ of {\it dark  matter}, and only about $5\%$ of {\it visible
matter}  in the Universe \cite{planck}.

Despite remarkable phenomenological achievements and many nice theoretical
properties, GR is not a complete
theory of gravity. It has well known long standing problems both in UV and IR
regimes. In UV or at short distances GR predicts singularities like the Big
Bang or black hole ones. In particular, under rather general
conditions, GR contains cosmological solutions which lead to an infinite matter
density at the beginning of the Universe \cite{bordeguthvilenkin}.
When  physical theory contains singularity, it is an evident indication  that
around  it such a theory has to be appropriately  modified. In IR pure GR does
not provide a decent explanation of the Dark Energy phenomenon since a possible
cosmological term has an unnatural and unexplainable tiny value.
From just a theoretical point of view GR is not a renormalizable theory even if
being quantized.
The ways to modify GR  usually  come from some more general theories like
quantum
gravity, string theory as well as astrophysical and
cosmological observations (for a review, see \cite{modreview}).
Unfortunately there is no so far solid fundamental physical principle which
could
tell
us how to find appropriate modification between infinitely many possible
theoretical constructions.

In the present paper we consider an analytic non-local modification of GR. This type of non-locality has
strong motivations from string field theory (SFT) \cite{ourreview} and $p$-adic
string theory \cite{padic}. The term ``analytic non-locality'' signifies that the theory contains non-local infinite derivative operators in the form of analytic functions of covariant derivatives. In particular SFT promotes analytic functions of the d'Alembert operator $\Box$.\footnote{That is, there is no a direct relation to another class of non-local gravity models based on the inverse of the d'Alembert operator \cite{woodard} or mixed models with positive and
negative powers of the d`Alembertian like in \cite{odin}.} Such a gravity modification was initially introduced in
\cite{Biswas:2005qr} and is intensively studied recently
\cite{Biswas:2011ar,Biswas:2012bp,koshelev,vernoviproc,Craps:2014wga,Modesto,dragovich-d,anupamnew}. One of the most interesting feature of analytic non-local gravities is the presence of a non-singular ghost-free bounce. The final stage of known bouncing solutions is a de Sitter expansion. The developed models contain a non-local term of the form $R\FF(\Box)R$ and it was shown already in \cite{Biswas:2012bp} that the late de Sitter phase after the bounce is stable.
In the present paper after highlighting the general ideas regarding this modified theory we also mainly
focus on the properties of a de Sitter (dS) solution.
We however get as the starting point a more general non-local term of the form
$$P(R)\FF(\Box)Q(R)\, .$$
The motivation for considering a more general non-local action is two-fold.
\begin{itemize}
	\item First, it follows from the above cited papers that the dependence on a non-local analog of $R^2$ covers all possible Lagrangians with respect to the dynamics around the dS space-time as long as those starting Lagrangians depend analytically on $R$, not only on derivatives. In our case we allow any, including non-analytic functions $P,Q$.
	\item Second, it is important to find out whether a pure dS background can be unstable in a non-local model. This will indicate a possibility to join bounce and inflation in one setup. Without such a joint behavior just an inflationary solution, namely the Starobinsky inflation \cite{Starobinsky:1980te}, was shown to successfully provide an inflationary background \cite{Craps:2014wga,Koshelev:2016xqb} in such non-local model of gravity with $P=Q=R$. Notice, that the Starobinsky solution is not preceded by a bounce. Also it is essential that a de Sitter phase of the inflation must be unstable. This guarantees the exit from inflation without which a matter creation is impossible.
\end{itemize}
Let us note here the results of \cite{anupamessay} in connection with the first point above. That paper is devoted to demonstrating an equivalence of quite a very general set of gravitational actions to just one action which containing exactly $P=Q=R$. We however emphasize that paper \cite{anupamessay} considers the equivalence w.r.t. quadratic variation of the action and the corresponding spectrum.
It does not touch in details the question of stability of classical perturbations and moreover those results do not account 3 and higher point correlations.
Thus apart from an attempt to seek for an unstable de Sitter background
the more and more active developments in measuring higher order correlations \cite{spider} will bring new constraints on models beyond the quadratic variation. An explicit computations of higher correlations is beyond the scope of the present paper but the formalism developed here is crucial in computations of those higher correlation functions.
Therefore
an altering of the initially proposed model may be required.

Before proceeding to the main part of our paper we would like to say that the non-local gravity models with analytic non-locality provide a modified Newtonian potential smoothing the singular limit at the origin. More specifically, the potential has a universal behavior which is a constant limit at zero distance for a very wide class of non-local functions entering the action while the standard $1/r$ falloff is naturally restored at large distances.
 Moreover various cosmologically interesting bounce solutions were build and thoroughly analyzed.
At the perturbative level an ability of such models to accommodate inflationary scenarios and in particular the Starobinsky inflation \cite{Starobinsky:1980te,Starobinsky:1982ee,Koshelev:2016xqb,Koshelev:2017tvv} was proven successful.
In particular, an embedding of the Starobinsky model in the non-local gravity leads to in principle testable modifications of the observable parameters, such that ration of the tensor and scalar power spectra $r$.
At the quantum level the non-locally modified gravity is shown to be renormalizable by the power-counting while the unitarity is preserved by construction. The latter means that there exist explicitly formulated conditions on non-local functions of the d'Alembert operator such that the spectrum of physical excitations is ghost-free moreover alongside with the renormalizability \cite{Stelle,Talaganis:2014ida} during the
quantization of the model.

The present paper is structured as follows. In Section~\ref{secmodel} we present our model and derive equations of motion. In Section~\ref{secsols}
we develop some general ideas of solution construction and discuss the most promising of
them.  In Section~\ref{secds} we turn to the de Sitter solution and formulate the
stability condition for linear perturbations. In Section~\ref{secgf} we derive explicitly the ghost-free condition for our model around a de Sitter background. In Section~\ref{secdssu} we explore the
stability condition for dS space-times in greater details and draw concluding remarks in
Section~\ref{seccodi}.

\section{Non-local generalization of GR and equations of motion}\label{secmodel}
We consider the following  nonlocal gravity action
\begin{equation} \label{lag:1}
S =  \int \dx\sqrt{-g}\left(\frac{M_P^2}2R-\Lambda + \frac{\lambda}2P(R)
\FF(\Box)Q(R) \right)\,  ,
\end{equation}
where $R$ is the scalar curvature, $\Lambda$ is the cosmological constant, $
\mathcal{F}(\Box)= \displaystyle \sum_{n =0}^{\infty} f_{n}\Box^{n}$  is an
analytic function  of the d'Alembert operator $\Box =
\nabla^\mu\nabla_\mu$ where $\nabla_\mu$ is the covariant derivative. The
Planck mass $M_P$ is related to the Newtonian constant $G$ as
$M_P^2=\frac1{8 \pi G}$ and $P$,$Q$ are scalar functions of the scalar
curvature.
$\lambda$ is a constant and in principle can be absorbed in the
rescaling of $\FF(\Box)$. However, it is a convenient tool to track the GR
limit which is $\lambda\to 0$.
As it is obvious we are going down the way of generalizing the results obtained
earlier in the case $P=Q=R$ \cite{Biswas:2012bp}. In the sequel we shall omit
an explicit further citation of these results referring them rather as
the \textit{non-local $R^2$} case.

To have physically meaningful expressions and to keep track of the
non-localities one should introduce the scale of non-locality using a new mass
parameter $\Mc$. Then the function $\FF$ would be expanded in Taylor series as
$\mathcal{F}(\Box)= \displaystyle \sum_{n =0}^{\infty}
f_{\Mc n}\Box^{n}/\Mc^{2n}$ with all constants $f_{\Mc n}$ dimensionless. We will return
to these notations during the discussion of our results.

Varying the action (\ref{lag:1}) with respect to the metric we get
the following equations of motion
\begin{equation} \begin{aligned} \label{EOMPQ}
		-\tilde G_{\mu\nu}\equiv&-M_P^2G_{\mu\nu}-g_{\mu\nu}\Lambda\\
&+\frac{\lambda}2g_{\mu\nu}P\FF(\Box) Q-\lambda(R_{\mu\nu}-K_{\mu\nu})V\\
&+\frac{\lambda}2\sum_{n=1}^\infty f_n\sum_{l=0}^{n-1}\left(P_\mu^{(\ell)}
Q_\nu^{(n-\ell-1)}+P_\nu^{(l)}
Q_\mu^{(n-l-1)}-g_{\mu\nu}(g^{\rho\sigma}P_\rho^{{(l)}}
Q_\sigma^{(n-l-1)}+P^{(l)} Q^{(n-l)})\right)=0\, .
\end{aligned} \end{equation}
Here $G_{\mu\nu}=R_{\mu\nu}-\frac12g_{\mu\nu}R$ is the Einstein tensor, $
K_{\mu\nu} = \nmu\nnu - g_{\mu\nu} \Box $,
 $V = P_R\FF(\Box)Q + Q_R\FF(\Box)P$ where the subscript $R$ indicates the
derivative w.r.t. $R$ (as many times as it is repeated) and
$$
P^{{(l)}}=\Box^l P,~P_\rho^{{(l)}}=\pd_{\rho}\Box^l P\text{ with the same
for }Q,~P_R,~\dots
$$
Provided there is a matter source as well the full equations of motion would contain $T_{\mu\nu}$ in the right hand side such that
\begin{equation}
	\tilde G_{\mu\nu}=T_{\mu\nu}
	\label{EOMPQmatter}
\end{equation}

Analyzing (\ref{EOMPQ}) we recognize that the first line is the canonical
EOM for the Einstein's
GR with the cosmological constant, the second line with $\FF(\Box)=1$
represents the extension to the local $f(R)$ type gravities while a
non-constant $\FF(\Box)$ as well as the last line are unique for a higher
derivative (probably non-local) modification of gravity.
The trace equation is of use and we write it separately
\begin{equation} \begin{aligned} \label{tracePQ}
&M_P^2R-4\Lambda + 2\lambda P \FF(\Box)Q -\lambda(R+3\Box)V -{\lambda}\sum_{n=1}^{\infty} f_n
\sum_{l=0}^{n-1}\left(g^{\rho\sigma}P_\rho^{{(l)}}
Q_\sigma^{(n-l-1)}+2P^{(l)} Q^{(n-l)}\right)   = -T^\mu_\mu.
\end{aligned} \end{equation}

If either $P$ or $Q$ is a constant in the action then effectively $\FF(\Box)$ trivializes to $f_0$ and we recover a local $f(R)$ gravity theory.
Note
that thanks to the integration by parts there is always the symmetry of an
exchange $P\leftrightarrow Q$.

\section{Solutions construction}\label{secsols}

\subsection{Cosmological FRW solutions}\label{subsecbi}

In this paper the primary goal is to attack the cosmological properties of the
proposed gravity model. This leads us to the cosmologically important metrics
from which we focus on the Friedmann-Robertson-Walker (FRW) configurations. The
latter have the following metric
\be
ds^2=-dt^2+a^2(t)\left(\frac{dr^2}{1-Kr^2}
+r^2d\theta^2+r^2\sin^2\theta d\varphi^2\right),\label{frwk}
\ee
where $t$ is the cosmic time, $a(t)$ is the scale factor and $K$ is the spatial
curvature. Such metrics represent a homogeneous and isotropic Universe. The corresponding tensor $\tilde G_{\mu\nu}$ in
(\ref{EOMPQ}) turns out to be diagonal with two distinct
components: $(00)$ and one of $(ii)$ for some spatial index $i=1,2,3$. It seems even simpler to work with the $(00)$ and the trace equations.
Hereafter we consider only the spatially-flat case $K=0$ unless explicitly stated otherwise.

The above chosen form of the metric implies that only the matter sources of the
form $T^\mu_\nu=\mathrm{diag}(-\rho,p,p,p)$ are compatible with the equations
of motion (\ref{EOMPQmatter}). Here $\rho$ is the energy density and $p$
is the pressure density. Such matter sources represent a wide class of
physically
important cases and include the most crucial set of perfect fluids which can be
written as $T_{\mu\nu}=(\rho+p)u_\mu u_\nu+pg_{\mu\nu}$ where $u_\mu$ is the
4-velocity of the fluid.

It is common that gravity is a theory with constraints thanks to Bianchi
identities. These identities manifest in the relation $\nabla^\mu
G_{\mu\nu}\equiv0$. However, one can check that given a diffemorphism invariant
action like as it is for (\ref{lag:1}) one ends up with a similar relation
$$\nabla^\mu
\tilde G_{\mu\nu}\equiv0\, .$$
This implies like in GR the canonical conservation equation for the matter
$$\nabla^\mu
T_{\mu\nu}=0\, .$$

In the sequel we will focus on special configurations which have the vanishing trace of Einstein equations. This can be either a vacuum such that $T_{\mu\nu}=0$ or radiation in case of a perfect fluid.
Indeed, radiation is characterized by $w=p/\rho=1/3$ and is therefore traceless.
In either situation solving the trace equation is almost enough. Namely, given that the trace is zero and there is a solution to the trace equation, and moreover accounting the FRW form of the metric we are left with just one equation, which we have chosen to be the $(00)$.
However, thanks to the Bianchi identity and the matter conservation this remaining equation can always be satisfied by adjusting the amount of radiation energy density.
To be physical, the radiation energy density must be positive.

\subsection{Non-local $R^2$ case}

It is clear from the complexity of expressions (\ref{EOMPQ}) that
constructing a general solution is a very ambitious hope. However we remind
that a significant progress has been achieved in the non-local $R^2$ case by
considering a simplifying ansatz
\begin{equation}\label{ansatzR}
\Box R=r_1R+r_2,
\end{equation}
with $r_{1,2}$ being
constants.
It is useful to repeat this procedure here in order to illuminate the steps
related to our model of interest (\ref{lag:1}).

Consider $P=Q=R$, the case originally discussed in \cite{Biswas:2005qr}. Then
equations (\ref{EOMPQmatter}) become
\begin{equation} \begin{aligned} \label{EOMPQR}
T_{\mu\nu}&=-M_P^2G_{\mu\nu}-g_{\mu\nu}\Lambda+\frac{\lambda}2g_{\mu\nu}
R\FF(\Box)
R-2\lambda(R_{\mu\nu}-K_{\mu\nu})\FF(\Box)R\\
&+\frac{\lambda}2\sum_{n=1}^\infty f_n\sum_{l=0}^{n-1}\left(R_\mu^{(l)}
R_\nu^{(n-l-1)}+R_\nu^{(l)}
R_\mu^{(n-l-1)}-g_{\mu\nu}(g^{\rho\sigma}R_\rho^{{(l)}}
R_\sigma^{(n-l-1)}+R^{(l)} R^{(n-l)})\right).
\end{aligned} \end{equation}
Application of ansatz (\ref{ansatzR}) means
\begin{equation}
\Box^nR=r_1^nR+r_2r_1^{n-1},~n>0\text{ and
}\FF(\Box)R=\FF_1R+\FF_2,
  \label{ansatzRn}
\end{equation}
where $\FF_1=\FF(r_1)$ and $\FF_2=\frac{r_2}{r_1}(\FF(r_1)-f_0)$.

As explained in the previous Subsection we will proceed with the trace
equation. For the traceless matter (or for no matter at all) it becomes
\begin{equation}
	A_1R-\lambda\FF^{(1)}(r_1)\left(2r_1R^2+ \partial_\mu R \partial^\mu
R\right)+A_2=0\, ,
\label{eqEinsteinRonlytraceansatz2}
\end{equation}
with
\begin{equation*}
\begin{aligned}
A_1&={}M_P^2-\lambda\left(4\FF'(r_1)r_2-2\frac{r_2}{r_1}
(\FF_1-f_0)+6\FF_1r_1\right)\, ,\\
A_2&={}-4\Lambda-\lambda\frac{r_2}{r_1}\left(2\FF'(r_1)r_2-2\frac{r_2}{r_1}
(\FF_1-f_0)+6\FF_1r_1\right)\, .
\end{aligned}
\end{equation*}
The above equation is satisfied provided $A_1=A_2=0$, and
$
\FF^{(1)}{(r_1)}=0
$. Here $\FF^{(1)}{(r_1)}$ is the first derivative w.r.t. the argument evaluated at point $r_1$.
Simple algebra gives
\begin{equation}
 r_2={}-\frac{r_1[M_P^2-6\lambda\FF_1r_1]}{2\lambda[\FF_1-f_0]}\, ,
\qquad
\Lambda={}-\frac{r_2M_P^2}{4r_1}\, . \label{r2lambda}
\end{equation}
Upon substitution of these relations in the $(00)$ equation one is left with the problem of positivity of a possible radiation
energy density as it is explained in the previous Subsection. The sign of $f_0$ would control the sign of the radiation energy density.

\subsection{Ansatz for equations of motion}\label{secsqrt}

Now we turn to our more generic model. To advance in the understanding of the structure of solutions we propose the following ansatz.
\begin{equation}
	P=Q\text{ and }\Box P=p_1P\, ,\label{ansatzP}
\end{equation}
where $p_{1}$ is a constant. Notice that for $P=R+\const$ ansatz (\ref{ansatzR}) is restored. Conditions (\ref{ansatzP}) provide dramatic simplifications.
Indeed, application of this ansatz yields
\begin{equation}
\Box^nP=p_1^nP\, ,~n>0\text{ and
}\FF(\Box)P=\FF_{p_1}P\, ,
  \label{ansatzRnp}
\end{equation}
where $\FF_{p_1}=\FF(p_1)$.
The further explicit substitution in the trace equation (\ref{tracePQ}) yields
(with $T^{\mu}_{\mu}=0$)
\begin{equation} \begin{aligned} \label{tracePQansatzp}
&M_P^2R-4\Lambda + 2\lambda \FF_{p_1}P^2
-2\lambda(R+3\Box)P_R\FF_{p_1}P  -{\lambda}\FF^{(1)}(p_1)(g^{\rho\sigma}P_\rho
P_\sigma+2p_1P^2)   = 0\, .
\end{aligned} \end{equation}

Analogously to the non-local $R^2$ case a condition $\FF^{(1)}(p_1)=0$
simplifies the clutter a lot and we are left just with the following equation
\begin{equation}
	M_P^2R-4\Lambda + 2\lambda \FF_{p_1}(P^2 -(R+3\Box)(P_RP))=0 \, .
	\label{localfreom}
\end{equation}
This equation is nothing but the trace equation of a local $f(R)$-type gravity with the following action
\begin{equation}
	S =  \int \dx\sqrt{-g}\left(\frac{M_P^2}2R-\Lambda + \frac{\lambda}2\FF_{p_1}P^2(R)\right)\, .
	\label{localfrs}
\end{equation}
As follows from general considerations in Subsection~\ref{subsecbi} the remaining $(00)$ equation evaluated on a solution to the trace equation can be satisfied by adjusting the radiation energy density. There are still enough parameters to control its sign.

Thus, we can draw the following conclusion: any solution of a local gravity of type (\ref{localfrs}) is a solution of our non-local theory as long as an additional condition (\ref{ansatzP}) is satisfied upon an appropriate adjustment of parameters in the non-local model. In particular, we must require $\FF^{(1)}(p_1)=0$. We note that for $P=R$ the trace equation coming from (\ref{localfrs}) and ansatz relation (\ref{ansatzP}) match upon adjustment of constant parameters as was shown in \cite{vernoviproc}. In a more general case the ansatz relation (\ref{ansatzP}) is indeed an extra condition and must be analyzed separately.

Then one can see that a specially simple case
arises if $P=\sqrt{R+R_0}$. For such a choice of function $P$ one gets
$P_R\sim1/P$ and as the result the latter equation can be solved by adjusting
the parameters. However, there is still a non-trivial equation to be satisfied
\be
\Box\sqrt{R+R_0}=p_1\sqrt{R+R_0}.\label{sqrt}
\ee
Nevertheless, presence of at least some solutions has been already preliminary shown in \cite{elena_in_progress}.
Interestingly, such a choice of $P$ has no  known local counterpart as it
would give just a canonical Einstein-Hilbert term with a cosmological constant.

\subsection{Constant curvature solutions}

This is a special class of solutions which include several important cases.

Substituting $R=\const$ into the trace equation
(\ref{tracePQ}) and using the fact that $P,Q,V$ are now constants one gets
\begin{equation} \begin{aligned} \label{tracePQdS}
&M_P^2R-4\Lambda + 2\lambda P f_0Q -\lambda RV
=0.\end{aligned} \end{equation}
Here $V$ reduces to $(P_RQ+Q_RP)f_0$.
Solving the latter equation is an algebraic rather than differential
problem.

The very important case is the de Sitter solution as far as it plays a
crucial role in the description of our Universe. In four dimensions it is
characterized by
\be
R_{\mu\nu}=\frac R4 g_{\mu\nu}\text{ and } R=\const>0.\label{dS}\ee
Notice that
just a constant $R$ does not mean the space-time
is de Sitter, however. One can check that dS space-time is a vacuum solution of our model,
i.e. no matter is needed to support it.

\subsection{Exact analytic bounce}

The following scale factor $a(t)$ in a spatially flat FLRW Universe results in an exact solution to full equations (\ref{EOMPQ})
\begin{equation}
	a=a_0\sqrt{\cosh(\sigma t)},
	\label{bouncePQ}
\end{equation}
At first one can find that this solution corresponds to $R=6\dot H+12 H^2=3\sigma^2=\const$ where as usual $H=\dot a/a$ and dot is the derivative w.r.t. the cosmic time $t$.
This effectively means that all the most non-trivial third line of equation (\ref{EOMPQ}) vanish.
As a consequence the trace of Einstein equations takes the form of (\ref{tracePQdS}) under an assumption that the matter is traceless. This equation in principle can be solved, at least by numeric methods. Recall that for a constant $R$ this is an algebraic equation.

The remaining equation is the (00) component of system (\ref{EOMPQ}). Using that $R_{00}=-3\dot H-3 H^2$ and computing explicitly $H$ this equation reads
\begin{equation}
	-M_P^2\frac R4(1-1/\cosh(\sigma t)^2)+\Lambda-\frac{\lambda}2Pf_0Q+\lambda V\frac R4(1+1/\cosh(\sigma t)^2)=T_{00}.
	\label{bouncePQ00}
\end{equation}
One immediately sees that all constant terms vanish thanks to equation (\ref{tracePQdS}) and one is just left with
\begin{equation}
	\frac R4\left(M_P^2+\lambda V\right)\frac 1{\cosh(\sigma t)^2}=T_{00}.	
	\label{bouncerad}
\end{equation}
As expected, this is the proper radiation energy density as it is proportional to $a^{-4}$. For a healthy model one must guarantee that the energy density is positive.

\section{Cosmological expansion}\label{secds}

\subsection{Background}

In this section we look for the cosmological spatially flat de Sitter which can
be written as
\be
ds^2=-dt^2+a_0^2e^{2Ht}d\vec x^2,
\label{dScosmo}
\ee
with a constant $H$, $t$ the cosmic time and the vector is the 3-dimensional
notion.
This is a particular case of a spatially flat FRW metric
\be
ds^2=-dt^2+a(t)^2d\vec x^2,
\label{FRW}
\ee
with $a(t)=a_0\exp(Ht)$. The general definition of $H$ is $H=\dot a/a$ with dot
being the derivative w.r.t. the cosmic time.

Some relevant background quantities (for a general $a$) are
\bea
R&=&12H^2+6\dot H , ~\Gamma_{ij}^0=H g_{ij}, ~\Gamma_{j0}^i=H\delta^i_{j},\nonumber\\
 \Box&=&-\pd_t^2-3H\pd_t+\frac{\delta^{ij}\pd_i\pd_j}{a^2},\nonumber
\eea
where the indexes $i,j$ range as $1,2,3$.
On the background all quantities are space homogeneous as the metric suggests.

For perturbations in many instances we employ
the conformal time $\tau$ such that
$$
ad\tau=dt.
$$
Then the general FRW metric
(\ref{FRW}) transforms to
\be
ds^2=a(\tau)^2(-d\tau^2+d\vec x^2).
\label{dScosmoconf}
\ee
For the de Sitter background (\ref{dScosmo})
$$
\tau=-\frac1{a_0H}e^{-Ht}~\Rightarrow~a(\tau)=-\frac1{
H\tau}.
$$
So when $t$ goes from past to future infinity, $\tau$ goes from $-\infty$ to
$0_-$. $t=0$ corresponds to $\tau=-\frac1{a_0H}$.

\subsection{Covariant perturbations}

Perturbations of equations (\ref{EOMPQ}) to the linear order around the dS
vacuum are easy to compute since many terms drop out. The variation of the metric is as usual
\begin{equation}
	g_{\mu\nu}=\bar g_{\mu\nu}+h_{\mu\nu}\, .
	\label{maindelta}
\end{equation}
Hereafter bars denote the background quantities.
What remains after a
careful computation is
\begin{equation} \begin{aligned} \label{deltaEOMPQ}
-&m^2\delta
G^\mu_{\nu}+(\bar R^\mu_{\nu}-\bar K^\mu_{\nu})v(\bar \Box)\delta R=0\, ,
\end{aligned} \end{equation}
where $m^2=M_P^2+\lambda f_0(\bar P_R\bar Q+\bar Q_R\bar P)$ and
$v(\bar \Box)=-\lambda(\bar P_{RR}\bar Q+\bar Q_{RR}\bar P)f_0-2\lambda \bar P_R\bar Q_R\FF(\bar \Box)$.
We have used the fact that variation of the $\Box$ acting on a scalar
function is a pure differential operator and all background curvatures are constants. Indeed,
varying the box we have
\be(\delta\Box)
f=[-h^{\mu\nu}(\pd_\mu\pd_\nu-\bar \Gamma^\rho_{\mu\nu}\pd_\rho)-\bar g^{\mu\nu}
\gamma^\rho_{\mu\nu}\pd_\rho]f.\ee
Here $\Gamma^\rho_{\mu\nu}$ is the Christoffel symbol.
Also we have used that starting from (\ref{maindelta}) one gets
\be
\delta g^{\mu\nu}=-h^{\mu\nu}
,~\delta \Gamma^\rho_{\mu\nu}=\gamma^\rho_{\mu\nu}=\frac12(\bar \nabla_\mu
h^\rho_\nu+\bar \nabla_\nu h^\rho_\mu-\bar g^{\rho\sigma}\bar \nabla_\sigma h_{\mu\nu}).
\ee
So if $f$ is a constant then $(\delta \Box) f=0$.
The same is true for $K^\mu_\nu$
\be
(\delta K^\mu_\nu)
f=[-h^{\mu\sigma}(\pd_\sigma\pd_\nu-\bar \Gamma^{\rho}_{\sigma\nu}
\pd_\rho)-\bar g^{\mu\sigma}\gamma^\rho_ { \sigma\nu }
\pd_\rho-\delta^\mu_\nu\delta\Box]f,
\ee
which is zero as long as $f$ is a constant.
Hence all the corresponding terms vanish.

Taking the trace of (\ref{deltaEOMPQ}) we get
\begin{equation} \begin{aligned} \label{deltatracePQ}
[m^2+(\bar R+3\bar \Box)v(\bar \Box)]\delta
R=\GG(\bar \Box)\delta R=0.
\end{aligned} \end{equation}
It is a homogeneous equation on $\delta R$. The general method of solving it
is to use the Weierstrass factorization
\be
\GG(\bar \Box)\delta R=-\prod_i(\bar \Box-\omega_i^2)e^{\gamma(\bar \Box)}\delta R=0\, ,\label{nlGeq}
\ee
where $\omega_i^2$ are the roots of the algebraic (or perhaps
transcendental) equation $\GG(\omega^2)=0$ and $\gamma(\bar \Box)$ is an entire
function (as a consequence $e^{\gamma(\omega^2)}$ has no roots).
We
assume that there are no multiple roots. Such roots complicate the story slightly but still can be treated analogously \cite{double}.
The overall minus in the LHS of the latter equation has no a particular meaning at this stage. However, it will precisely match to the minus sign upon derivation of the ghost-free condition for scalar modes below in Section~\ref{secgf}.
Then we can solve (\ref{nlGeq}) for each $\omega_i$ separately
\be \label{eqomega}
(\bar \Box-\omega_i^2)\delta R=0.
\ee
The latter is just a second order linear differential equation.
It can be written
explicitly as
\be
\left(\pd_\tau^2-\frac2\tau\pd_\tau+{k^2}+\frac{\omega_i^2}{H^2\tau^2}
\right)\delta R=0,
\label{eqdRdS}
\ee
where we have taken the dS form of the background.
The solution yields
\be
\delta R_i=(-k\tau)^{ 3/2} \left( C_{1i} J_{\nu_i}(-k\tau) + C_{2i}
Y_{\nu_i}(-k\tau)\right),
\label{dRdS}
\ee
where $J,~Y$ are the Bessel functions
of the first and second kinds,
respectively, with $\nu_i=\sqrt{\frac94-\frac{\omega_i^2}{H^2}}$ and
$C_{1i,2i}$ are the integration constants.

For small values of $\tau$ which correspond to large cosmic times $t$ the Bessel functions have the following asymptotic
behavior
\bea
J_\nu(z) &\sim& z^{\Re\nu}\, ,\nonumber\\
Y_\nu(z) &\sim& z^{-|\Re\nu|} \, \text{ for }\Re\nu\neq 0\, ,\nonumber\\
Y_\nu(z) &\sim& \ln z  \, \text{ for }\Re \nu = 0\, .\nonumber
\eea
From this we conclude that $\delta R_i$ are bounded irrespectively of
the boundary conditions provided
\begin{equation}
  |\Re\nu_i|<\frac32.
  \label{condnu}
\end{equation}

The full answer for $\delta R$ is
\be
\delta R=\sum_i\delta R_i.
\label{dRdSfull}
\ee
where each $\delta R_i$ has its arbitrary integration constants. The only thing
we care about is that the total $\delta R$ must be real.
Note that
$\delta R=0$ is an always existing and not necessarily trivial configuration.

\subsection{Scalar perturbations}

We are focusing on scalar classical perturbations only since the behavior of vector and tensor classical perturbations remain the same as in GR. The corresponding equations are just scaled by a constant factor thus not affecting the dynamics at all. This is because the introduced gravity modification does not awake neither scalars nor vectors. This will become even more apparent in the next Section where the spin-2 and spin-0 modes quadratic actions around dS will be computed explicitly.

The metric for the scalar perturbations around a FRW background is defined as
\begin{equation}
ds^2=a(\tau)^2\left[-(1+2\phi)d\tau^2-2\pd_i \beta d\tau
dx^i+((1-2\psi)\delta_{ij}+2\pd_i\pd_j\gamma)dx^idx^j\right].
\label{alex_scalar_pert}
\end{equation}
Here we employ the standard ADM decomposition and the notions of spin are w.r.t. the spatial part of the metric.
From 4 scalar modes only 2 are gauge invariant.
The convenient gauge invariant variables (Bardeen potentials) are introduced as
\begin{equation}
\Phi=\phi-\frac{1}{a}(a\vartheta)^\prime=\phi-\dot
\chi,\qquad\Psi=\psi+\Hc\vartheta=\psi+H\chi,
\label{GIvars}
\end{equation}
where  $\chi=a\beta+a^2\dot\gamma$, $\vartheta=\beta+\gamma'$, $\Hc(\tau)=a'/a$.
The prime denotes the differentiation with respect to the conformal time
$\tau$ and the dot as before w.r.t. the cosmic time $t$.

The $(1+3)$ structure suggests
to represent the perturbation quantities (which can depend on all 4
coordinates) as
\be
f(\tau, \vec x)=f(\tau,k)Y(k,\vec x),
\label{scalar13}
\ee
where $k=|\vec k|$ comes from the definition of the $Y$-functions as spatial
Fourier modes
\be
\delta^{ij}\pd_i\pd_j Y=-k^2Y.
\ee
Obviously
\be
Y=Y_0e^{\pm i\vec k\vec x}.
\ee
The relevant expressions for the background d'Alembertian is
\be\Box=-\frac{1}{a^2}\pd_\tau^2-2\frac{a'}{a^3}\pd_\tau-\frac{k^2}{a^2}
=-\partial_t^2-3H\partial_t-\frac{k^2}{a^2}.
\ee
Notice, that
all the above expressions in this subsection are valid for a generic scale
factor $a$.


What we want to determine are the Bardeen potentials introduced in
\eqref{GIvars}. To do this we need two equations.

One equation is given by the
formulation of $\delta R$ in terms of $\Phi$ and $\Psi$ accounting that
the time behavior of $\delta R$ itself is found above. A lengthy computation
gives
\begin{equation}
	\widetilde{\delta R}= \frac{2}{a^2}\left[k^2
(\Phi-2\Psi)-3\frac{a'}{a}\Phi'-6\frac{a''}{a}\Phi-3\Psi''-9\frac{a'}{a}
\Psi'\right],
\label{deltaRGI}
\end{equation}
where $\widetilde{\delta}R=\delta R- \bar R'(\beta+\gamma')$ is a convenient gauge invariant analog of $\delta R$.
This expression is valid for a general $a$ while its dS form is
\begin{equation}
	\widetilde{\delta R} = -6H^2 \left(4\Phi - \tau(\Phi' + 3\Psi') + \tau^2 \Psi''\right)
+ 2\tau^2 H^2 k^2 \left(\Phi -2\Psi\right).
\label{deltaRGIdS}
\end{equation}

Another equation can be
got using the following procedure. Let us exploit all the system of equation (\ref{deltaEOMPQ}).
First we derive the $i\neq j$ component of the system (\ref{deltaEOMPQ}) which reads
\begin{equation} \begin{aligned} \label{deltaEOMPQijfinal}
		-&m^2(\Phi-\Psi)+v(\bar \Box)\widetilde{\delta R}=0.
\end{aligned} \end{equation}
Second we write down the $(0i)$ equation of the system (\ref{deltaEOMPQ}) which is
\begin{equation} \begin{aligned} \label{deltaEOMPQ0jfinal}
		&2m^2(\Psi'+\Hc\Phi)+(v(\bar \Box)\widetilde{\delta R})'-\Hc v(\bar \Box)\widetilde{\delta R}=0.
\end{aligned} \end{equation}
Third, we deduce the (00) equation of system (\ref{deltaEOMPQ}) which yields
\begin{equation} \begin{aligned} \label{deltaEOMPQ00final}
		-&2m^2(k^2\Psi +3\Hc\Psi'+3\Hc^2\Phi)-3\Hc (v(\bar \Box)\widetilde{\delta R})'-\left(k^2-\frac3{\tau^2}\right)v(\bar \Box)\widetilde{\delta R}=0,
\end{aligned} \end{equation}
where the last term proportional to $1/\tau^2$ accounts the fact that the background is a dS space.
Finally, multiplying (\ref{deltaEOMPQijfinal}) by $k^2$, (\ref{deltaEOMPQ0jfinal}) by $3\Hc$ and summing these results all together with (\ref{deltaEOMPQ00final}) as well as accounting that for the dS space-time $\Hc^2=1/\tau^2$ we get
\begin{equation}
	-m^2k^2(\Phi+\Psi)=0,
	\label{constraint}
\end{equation}
which is our required another equation. Obviously the latter constraint simplifies the succeeding computations considerably.
Since $\widetilde{\delta R}$ is provided independently by (\ref{dRdSfull}) one
readily expresses from (\ref{deltaEOMPQijfinal})
\be
2\Phi=\frac1{m^2}\sum_i v(\omega_i^2)\delta R_i.\label{phiminuspsisol}
\ee
Notice that $\widetilde{\delta R}$ coincides with $\delta R$ if $\bar R$ is a constant or on a more general basis in the longitudinal (Newtonian) gauge $\beta=\gamma=0$.
Taking into account (\ref{condnu}) we see that Bardeen potentials are vanishing as long as $|\Re \nu_i|<3/2$ for each $i$. If $\Re \nu_i=0$ for some $i$ then the perturbations become frozen. At last, if for at least one $i$ we have $|\Re\nu_i|>3/2$ then perturbations grow.

This is in perfect agreement with \cite{Biswas:2012bp} and the comparison should
be done as follows. In that reference a more general class of solutions was
studied
which asymptote to the de Sitter background at late times while the non-local
part of the Lagrangian is given by
$R\FF(\Box)R$.
In the present paper we stick to the de Sitter solutions from the very
beginning but the non-local part of the Lagrangian is more general. Assuming in
our case $P(R)=Q(R)=R$ we retrieve the results exactly as in Section~4 in
\cite{Biswas:2012bp} where the late time regime is studied.

There is a special case $m^2=0$ as we loose the possibility to find out the
Bardeen
potentials separately. The compatibility of system
(\ref{deltatracePQ},\ref{deltaEOMPQijfinal}) requires that either $\delta R=0$
or there is
a root of $v(\Box)$. However, neither of equations coming from
(\ref{deltaEOMPQ}) would help in this case as all of them lack information
about individual Bardeen potentials if $m^2=0$.
Physically this reflects the fact that effectively the Einstein-Hilbert term
vanishes and one reduces the number of propagating degrees of freedom.

\section{No-ghost conditions}\label{secgf}
\subsection{Physical excitations}

In this Subsection we find out the physical excitations and their respective quadratic Lagrangians around dS background for the model (\ref{lag:1}). dS is a maximally symmetric space and this helps a lot in technical calculations. This can be accomplished in a fully covariant way using (\ref{maindelta}).
To work out the task we use the covariant mode decomposition introduced as \cite{decomp}
\begin{equation}
	h_{\mu\nu}=\tilde h_{\mu\nu}+(\bar\nabla_\mu A_\nu+\bar\nabla_\nu A_\mu)+\left(\bar\nabla_\mu\bar\nabla_\nu-\frac{\bar g_{\mu\nu}}4\bar \Box\right)B+\bar g_{\mu\nu}\frac h4.
	\label{decomp}
\end{equation}
Here $\tilde h_{\mu\nu}$ is a transverse and traceless tensor (spin-2), $A_\mu$ is a transverse vector (spin-1), $B$ and $h$ are scalars (spin-0) and the operator acting on $B$ is traceless. Notice that here we use the fully covariant 4-dimensional decomposition and notions of spin are w.r.t. the full 4-dimensional Poincar\'e group. From a pure group representation arguments modes of different spins do not mix at the linearized level and therefore we can analyze them separately.

Let us first introduce few intermediate notations:
\begin{eqnarray}
	\sqrt{-\bar g}\delta_{EH}&=& \delta^2\left(\sqrt{-g} R\right), \, \,   \delta_g=\frac{h^2}8-\frac{h_{\mu\nu}h^{\mu\nu}}4,\nonumber
\end{eqnarray}
where specifically around the dS background we have
\begin{equation}
\begin{split}
	\delta_{EH}&=\left(\frac14 h_{\mu\nu}\bar\Box
h^{\mu\nu}-\frac14h\bar \Box h+\frac12h\bar\nabla_\mu\bar\nabla_\rho
h^{\mu\rho}+\frac12\bar\nabla_\mu h^{\mu\rho}\bar\nabla_\nu h^\nu_\rho\right)-\frac{\bar R}{24}(4h_{\mu\nu}^2-h^2).
\end{split}
\label{delta0lambdads}
\end{equation}
This can be checked using the results of \cite{NPB}. We also need a variation of the scalar curvature which is
\begin{equation}
	\delta R=(-\bar R_{\mu\nu}+\bar\nabla_\mu\bar\nabla_\nu-\bar g_{\mu\nu}\bar\Box)h^{\mu\nu}.
	\label{deltarmunu}
\end{equation}
This expression above is valid for any background.

We are now ready to compute the second variation of the full action (\ref{lag:1}) around dS background. To start with we note that this task for $P=Q=R$ was accomplished already in \cite{BKMessay,BKMfull}. Generically this is a very tough task but the constancy of the background scalar curvature helps a lot. It is performed by considering $\FF(\Box)$ as a Taylor series and performing a number of summations, integrations by parts and algebraic simplifications as explained in the just mentioned references. Essentially here we just need to generalize to arbitrary $P$ and $Q$. To simplify the things we are using the background equation (\ref{tracePQdS}) wherever possible. The result of these considerations is
\begin{equation}
\begin{split}
	\delta^2S=\int d^4x\sqrt{-\bar g}&\left[\left(\frac{M_P^2}2+\frac\lambda2(P_RQ+Q_RP)f_0\right)\left(\delta_{EH}-\frac {\bar R}2\delta_g\right)+\right.\\
	&+\left.\frac\lambda2\frac12(P_{RR}Q+PQ_{RR})f_0(\delta R)^2+\frac\lambda2P_RQ_R\delta R\FF(\bar\Box)\delta R\right],
\end{split}
	\label{d2actionds}
\end{equation}
where all $P,Q$ quantities are evaluated on the solution to equation (\ref{tracePQdS}). Note also that $\Lambda$ is not explicitly here due to the use of relations (\ref{r2lambda}).

It is very important for the subsequent computation that around the dS background the terms in $h_{\mu\nu}$ which contain $A_\mu$ and $\bar\nabla_\mu\bar\nabla_\nu B$ do not contribute neither to $\delta_{EH}-\frac{\bar R}2\delta_g$ nor to  $\delta R$.
This implies that formula (\ref{decomp}) reduces around the dS space to
\begin{equation}
	h_{\mu\nu}=\tilde h_{\mu\nu}+\frac14g_{\mu\nu}\phi,\quad \phi=-\Box B +h.
	\label{decompds}
\end{equation}
This looks like a tremendous simplification.

The spin-2 excitation $\tilde h_{\mu\nu}$ does not appear in the linearization of the non-local term $P\FF Q$ apart from the constant term in the Taylor series expansion of $\FF$. Simply $\delta R$ evaluated on $h_{\mu\nu}=\tilde h_{\mu\nu}$ vanish. Therefore the spin-2 propagator remains like in a local $R^2$ gravity. Its careful evaluation reveals the following result
\begin{equation}
	\delta^2 S(\hat h_{\mu\nu})=\int d^4x\sqrt{-\bar g}\frac12\hat{\tilde h}_{\mu\nu}\left(\bar \Box-\frac {\bar R}6\right)\left[1+\frac2{M_P^2}\frac\lambda2 f_0(PQ)_R\right]\hat{\tilde h}^{\mu\nu},
	\label{spin2prop}
\end{equation}
where $f_0$ is the zero Taylor coefficient the $\FF(\Box)$ decomposition and the hat signifies that we have canonically normalized the field (i.e. we would have $\frac12\hat{\tilde h}_{\mu\nu}\Box\hat{\tilde h}^{\mu\nu}$ as the dynamical term for the field in GR). We notice that the expression in brackets is a purely background quantity and is therefore a constant. We just need to guarantee it is positive to preserve the no-ghost spectrum.

The spin-0 mode $\phi$ is a bit less trivial as it does contribute to the variations of the non-local term, which is a natural expectation. $\delta R$ evaluated on $h_{\mu\nu}=\frac14\bar g_{\mu\nu}\phi$ is just $-\frac14(3\bar \Box+\bar R)\phi$. One readily arrives at
\begin{equation}
	\begin{split}
		\delta^2 S(\phi)&=-\int d^4x\sqrt{-\bar g}\frac{\hat{\phi}}2\left(\bar \Box+\frac {\bar R}3\right)\left[1+\frac\lambda{M_P^2} f_0(PQ)_R-\right.\left.\frac{\lambda}{M_P^2}\left(f_0(P_{RR}Q+PQ_{RR})+2P_RQ_R\FF(\bar\Box)\right)(3\bar\Box+\bar R)
	\right]\hat\phi.
	\end{split}
	\label{spin0prop}
\end{equation}
where as before the hats denote a canonically normalized field.

\subsection{Notes on spin-$0,2$ modes dynamics}

We see that the derived above result for the Lagrangians of scalar and tensor modes have a very neat structure. Namely, they clearly resemble the canonical factors known from GR (or from local $R^2$ modification of gravity) but on top of this the quadratic form of the scalar mode has a new non-local components.

First we stress that in order to preserve the nature of the tensorial mode we must arrange that
\begin{equation}
	\Tc=1+\frac{\lambda}{M_P^2}f_0(PQ)_R>0.\label{no2ghost}
\end{equation}
This impose certain restrictions on functions $P,Q$ as well as signs of parameters $\lambda$ and $f_0$. For example we notice that for $P=1/Q$ the above combination is always 1, i.e. greater than zero. This can effortless be generalized to $PQ=\mathrm{const}$ configuration. In order to connect this result with the previous Section we note that the 3-dimensional spin-2 excitations form a subset of the 4-dimensional spin-2 modes. This is why the dynamics of tensor perturbations in the ADM formalism does not change in our model.

The situation is more involved for the scalar mode. The first factor $\bar\Box+\bar R/3$ corresponds to the standard GR scalar degree of freedom. The second factor can generate new degrees of freedom. As many as  roots it has w.r.t. the d'Alembert operator. Given $\FF(\Box)=f_0$ we observe the standard Brans-Dicke scalar mode of the $f(R)$ gravity. In a general situation with an arbitrary (analytic) function $\FF(\Box)$ we must require that no more than one root exist.

Having said this we formalize the statement as follows. We demand that the following representation takes place:
\begin{equation}
\begin{split}
	&1+\frac2{M_P^2}\frac\lambda2 f_0(PQ)_R-\frac{\lambda}{M_P^2}f_0(P_{RR}Q+PQ_{RR})(3\bar\Box+\bar R)-\frac{\lambda}{M_P^2}2P_RQ_R(3\bar \Box+\bar R)\FF(\bar\Box)=-(\sigma\bar\Box-\omega^2)e^{\gamma(\bar\Box)},
\end{split}
	\label{spin0roots}
\end{equation}
with $\gamma(\Box)$ being an entire function resulting in no roots from the exponential factor.
We observe at this stage that the LHS of the latter equation is exactly $\Gc(\bar\Box)/M_P^2$ where the operator $\Gc(\Box)$ is defined in (\ref{deltatracePQ}).
The minus sign on the right hand side accounts for the overall minus sign in (\ref{spin0prop}). This means that for $\sigma=1$ and $\omega^2>0$ we have a good residue sign $+1$ in this new pole and keep the wrong residue sign $-1$ for the standard gravity scalar (the latter is required to arrange the proper IR behavior of the full graviton propagator). Hence, we demand that $\sigma=0$ (no roots) or $\sigma=1$ when the Brans-Dicke scalar appears. $\omega^2$ is a real positive mass squared of the scalar mode so that it is not a tachyon.

In order to satisfy the above relation one must adjust correspondingly function $\FF(\Box)$. The initial assumption that $\FF(\Box)$ is analytic is obviously fulfilled. To gain more constraints we can check for example the consistency of the above expression at $\bar\Box=0$. Indeed, evaluating at zero one gets
\begin{equation*}
	1+\frac{\lambda}{M_P^2}f_0(PQ)_R-\frac{\lambda}{M_P^2}f_0\bar R(PQ)_{RR}=\omega^2e^{\gamma(0)}>0.
\end{equation*}
First we notice that in analogy with the spin-2 excitation given $PQ=\mathrm{const}$ the latter inequality is satisfied.

However, a bit more general consideration comes from considering the latter inequality as a differential equation w.r.t. $PQ$ as the function of $\bar R$. Generically we must analyze all possible classes of functions which may satisfy the inequality. This will be hopefully accomplished in a separate paper in conjunction with many other open questions related to the advertised here model. Nevertheless, we want to point out the following two interesting configurations.
\begin{itemize}
	\item Requiring that $(PQ)_R-\bar R (PQ)_{RR}$ is a constant we immediately see that $(PQ)_R=2\alpha\bar R+\beta$. Consequently, $PQ=\alpha\bar R^2+\beta\bar R+\gamma$ favoring generalized $R^2$ models of the form $R\FF R$ provided $\alpha\neq0$.
	\item Second and even more intriguing case is when $(PQ)_R=\beta$. This is in a sense a sub-case of the above more general formula but it has a very interesting interplay with the whole model. First, this results in $(PQ)_{RR}=0$ and a very simple constraint on parameters. However, on top of  this special constraint results in $PQ=\beta\bar R+\gamma$. This in turn accommodates explicitly a possibility $P=Q=P_0\sqrt{R+R_0}$. Overall this looks like a de-localized EH gravity. The main point to make here though is that as we have seen above in Section~\ref{secsqrt} such a peculiar form of $P,Q$ functions leads to a possibility to solve the full Einstein equations in this model. An extended analysis of this special configuration will be undertaken in a future publication \cite{brankonext}.
\end{itemize}

\section{Stability of the constant curvature backgrounds}\label{secdssu}

As the main result of Section~\ref{secds} the question of stability of the de
Sitter vacuum is narrowed to the satisfactory solution to eq. (\ref{condnu}).
$\nu$ in turn depends on the structure of the non-local operator $\GG(\Box)$
such that
\be
\nu=\sqrt{\frac94-\frac{\omega^2}{H^2}},\label{nunu}
\ee
and $\GG(\omega^2)=0$.

Following equation (\ref{spin0roots}) and the comment thereafter we have an explicit form for the operator $\GG$.
Moreover, as we have already understood in the previous Section the requirement that the system does not lead to ghosts demands that
there is no more than one such a root $\omega^2$ for the operator
$\GG$.

To make the analysis of roots tractable we specialize to monomials \cite{ivan}
\be
P(R) = R^p, \quad Q(R)=R^q,
\ee
for some nonzero $p$ and $q$.
At the background level this results in the following modification of equation
(\ref{tracePQdS})
\begin{equation} \begin{aligned} \label{tracePQdSpq}
&M_P^2R-4\Lambda + \lambda f_0R^{p+q}(2-p-q)
=0.\end{aligned}\end{equation}
This equation can be solved in general w.r.t. $R$ as long as $p,~q$ are integers and $-3\leq p+q\leq4$.
As in the previous Section it is useful to analyze
$\GG(0)$ to see whether the
stability condition can be reached. Indeed, $\GG$ is analytic by
construction but a compatibility condition must be fulfilled
\be
M_P^2+\lambda R^{p+q-1}(p+q)(2-p-q)f_0=\omega^2e^{\gamma(0)}.
\label{pqn0}
\ee
It is obvious from (\ref{nunu}) that as long as $\omega^2$ is real it should be
at least positive in order to satisfy (\ref{condnu}). The latter condition
(\ref{pqn0}) clearly shows that $\omega^2$ is indeed real and therefore reduces
to the following necessary inequality
\be
M_P^2+\lambda R^{p+q-1}(p+q)(2-p-q)f_0>0.
\label{pqn01}
\ee
Satisfactory solution to this relation is a necessary stability condition.
From here it is obvious that two special cases, namely $p+q=0$ and $p+q=2$
always have a stable de Sitter phase.

In a general situation we have to understand equation
(\ref{tracePQdSpq}) together with the latter inequality (\ref{pqn01}).
As a side turn one can simplify {(\ref{pqn01})} using
the background equation
(\ref{tracePQdSpq}) to
\be
M_P^2R(p+q-1)<{4\Lambda}(p+q).
\ee
One can see from this representation that for $p+q=1$ (directly related to
(\ref{sqrt})) one can have a stable de Sitter phase given a positive $\Lambda$.
The latter condition is not improbable as long as $\lambda f_0>0$.

In an attempt to solve the system \eqref{tracePQdSpq} and \eqref{pqn01}
one can rewrite it as
\begin{equation}\label{simple}
  1 -s + u = 0, \qquad 1+ u z >0,
\end{equation}
where $s = \frac{4\Lambda}{M_P^2R}, \, z=p+q, \, u = \frac{\lambda f_0}{M_P^2}
R^{z-1}(2-z)$. This latter system looks rather simple but unfortunately does
not provide immediate new interesting solutions from the physical point of view.

\section{Discussion and conclusion}\label{seccodi}

The main subject of our investigation was the generalized gravity theory
represented by (\ref{lag:1}) with the emphasis on the de Sitter solution. This
theory contains higher derivatives and as such has a potentially dangerous
property to generate ghosts. However, the ghost-free condition can be clearly
formulated and is translated to the following statements: (i) the operator $\GG(\Box)$ (eq.
(\ref{deltatracePQ}) or explicitly the LHS of equation (\ref{spin0roots}) multiplied by $M_P^2$) has no more than one root; (ii) one should arrange $\sigma=0$ or 1 and $\omega^2$ is positive in (\ref{spin0roots}); (iii) $\Tc$ in (\ref{no2ghost}) must be positive. Going further we find that
simple algebraic conditions \eqref{tracePQdSpq} and \eqref{pqn01} determine
a possibility to have stable dS configurations.

Namely, satisfying the inequality
\eqref{pqn01} one guarantees that the de Sitter phase is stable.
Interestingly, for $P=R^p,Q=R^q$ two special cases $p+q=0$ and $p+q=2$ always lead to
stable de Sitter solutions. The first case is a generalized ``cosmological
constant'' (since $PQ=1$).
The second case is the generalized $R^2$ theory and for
now is the mainly developed candidate for a renormalizable theory of gravity
\cite{Talaganis:2014ida}.

Generically one would expect such a non-local model to be a candidate for the
UV gravity completion. In this case a possible de Sitter phase would be
attributed to the inflationary stage of the Universe evolution. This way the de
Sitter phase should not be eternally stable as the inflation should stop. This
implies for example that the presently found bounce solutions in non-local gravity with $P=Q=R$ \cite{Biswas:2012bp,mevernov} cannot be straightforwardly continued to a viable cosmological inflation. This is because those solutions end up with a dS expansion which is eternally stable as it corresponds to $p+q=2$ in (\ref{pqn01}). Even though the Starobinsky inflation is here, in the model with $P=Q=R$, this solution has no a bounce phase while an idea to construct a joint bounce and inflation is an intriguing open question on the table.

On contrary one can try to use non-local models in an attempt to challenge the
present day slowly accelerated expansion of the Universe. In this case one has
to arrange a
stable evolution for a long period of time to be compatible with the presently
observed Universe. In this regime models with $P=Q=1$ seem to be absolutely suitable.

One more observed special case is a generalized Einstein-Hilbert term with
$p=q=1/2$. This leads to a possibility to solve completely the equations of
motion provided we can find a solution to equation (\ref{sqrt}). This solution will automatically be a solution of a local gravity action (\ref{localfrs}). However,
finding a solution to (\ref{sqrt}) turns out to be a complicated problem and we keep it in a
separate study \cite{brankonext}.

Generically presence or absence of stable de Sitter configurations in model
(\ref{lag:1}) with monomial $P$ and $Q$ has been reduced to a really simple
algebraic system (\ref{simple}). In a general situation both stable and
unstable regimes can be organized based on the parameters of the model.
In all regimes we can control the order of the non-locality
scale $\Mc$ using other parameters in the theory.

Some other cosmological solutions of modified gravity with analytic nonlocality
can be found in \cite{sols}.


\end{document}